\documentclass[11pt]{article}
\usepackage{moriond,epsfig}

\def\be{\begin{equation}}
\def\ee{\end{equation}}
\def\bea{\begin{eqnarray}}
\def\eea{\end{eqnarray}}

\newcommand{\SuperK}    {Super--Kamiokande}
\def\Cherenkov         {Cherenkov\ }
\newlength{\figwid}
\begin{document}
\vspace*{4cm}
\title{Latest Results from the K2K experiment\footnote{Talk presented at the XXXVIIth 
 Rencontres de Moriond "Electroweak Interactions and Unified Theories",
 Les Arcs, Savoie, France, March 9-16, 2002.}}

\author{Atsuko K. Ichikawa\\(for the K2K Collaboration)}

\address{Institute of Particle and Nuclear Studies, KEK,
  Tsukuba, Japan}

\maketitle\abstracts{
The KEK to Kamioka long-baseline neutrino experiment (K2K) is 
the first accelerator-based experiment with hundreds of  km neutrino 
path length.
K2K focuses on the study of the existence of 
neutrino oscillations in $\nu_{\mu}$ disappearance 
that is observed in atmospheric neutrinos.
With nearly pure muon neutrino beam produced
with $5.6\times 10^{19}$ protons from 12 GeV PS on target,
56 neutrino events have been observed at Super-Kamiokande,
the far detector at 250 km distance.
The expectation is $80.6^{+7.3}_{-8.0}$ derived from the measurement
at the near site.
The experiment expects to accumulate $10^{20}$ protons on
target, providing sufficient statistics to study neutrino oscillations by
spectral analysis for $\nu_{\mu}$ disappearance.
}

\section{Introduction}
\indent
The existence of neutrino oscillations implies that neutrinos
are massive and that flavor and mass eigenstates are 
mixed in the lepton sector, too.
Experiments on atmospheric neutrinos have found a significant 
deficit in the flux of $\nu_{\mu}$ which have travelled an earth
scale distance\cite{deficitALL:ref,subGeV:ref,multi:ref}.
The interpretation of these results provides not only strong evidence
of $\nu_{\mu}\rightarrow 
\nu_{\tau}$ (or $\nu_{\mu}\rightarrow \nu_{\mathrm sterile}$)
oscillations\cite{SKatm:ref,SKster:ref} but also evidence for 
different mixing properties in the lepton and quark sectors.
 For a neutrino energy $E_{\nu}\rm (GeV)$ and a distance from the 
source $L\rm (km)$, the oscillation probability can be written in 
terms of the mixing 
angle $\theta$
and the
difference of the squared masses $\Delta m^2$ (eV$^2$) in two flavor
approximation as
\mbox{
$P(\nu_{\mu}\rightarrow\nu_x)=\sin^22\theta\sin^2 (1.27\Delta m^2 L/E_{\nu})$.
}

The KEK to Kamioka long-baseline neutrino experiment (K2K) is
the first accelerator-based
experiment with hundreds of  km neutrino path length\cite{K2Kletter}.
The intense, nearly pure neutrino beam 
(98.2\% $\nu_{\mu}$, 1.3\% $\nu_e$, and 0.5\% $\overline{\nu}_{\mu}$)
has an average $L/E_\nu\approx\!200 (L=250 {\rm km},
                   \langle E_\nu\rangle\approx\!1.3 {\rm GeV})$.
The neutrino beam properties are measured just after production,
and the kinematics of parent pions are measured {\em in situ\/} to
extrapolate  the measurements 
at the near detector to the expectation at the far detector.
  K2K focuses on the    study     of the existence of 
neutrino oscillations in $\nu_{\mu}$ disappearance that is
observed in atmospheric neutrinos, and on the search for 
$\nu_{\mu}\rightarrow \nu_e$ oscillation with well understood flux and
neutrino composition in the $\Delta m^2 \ge  
                               2\times 10^{-3}\rm eV^2$ region.
In this paper, the results on the 
event rates with data taken from June 1999
to July 2001 corresponding to $5.6\times 10^{19}$ 
protons on target are described.

\section{Experimental Setup and Analysis}

\begin{figure}[htbp]
  \centerline{\psfig{file=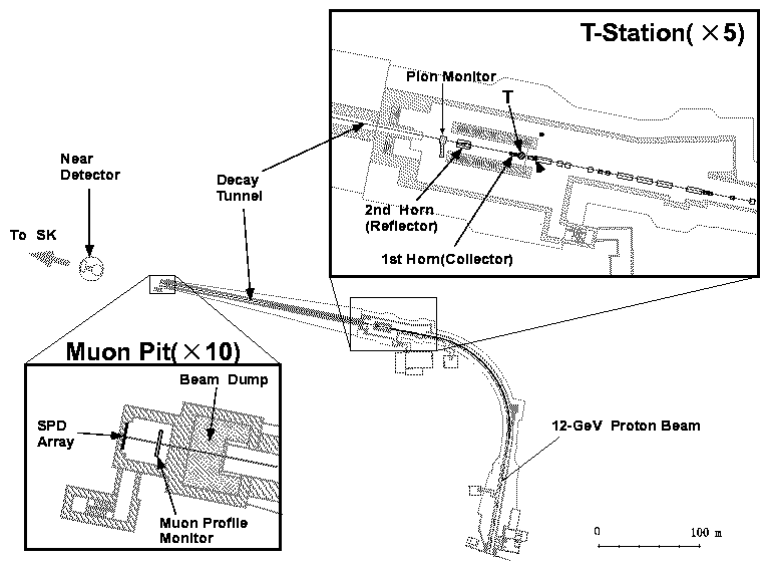,width=\figwid}}
  \caption{Near site setup at KEK.}
\label{fig:beamline}
\end{figure}

The K2K neutrino beam\cite{align:ref,beamline:ref} is a horn-focused 
wide band $\nu_{\mu}$ beam.
Figure~\ref{fig:beamline} shows the setup at KEK site.
The primary beam for K2K is 12 GeV kinetic energy 
protons from the KEK proton-synchrotron (KEK-PS)\cite{PS:ref}.
Every 2.2 s, approximately $6\times 10^{12}$ protons in nine bunches are
fast-extracted in a single turn, making a 1.1 $\rm\mu{s}$ beam spill.
These protons are focused onto a 30 mm diameter, 66 cm long
aluminum target which
is a current carrying element in the first of a pair of horn
magnets operating at 250 kA\cite{horn:ref}.
This design maximizes the efficiency of the toroidal magnetic field to focus 
positive pions produced in proton-aluminum interactions,
while sweeping out negative secondary particles.

Downstream of the horn system, before the 200 m long decay volume 
where the pions decay to $\nu_\mu$ and muons, 
a gas-\Cherenkov detector 
(\mbox{PIMON})\cite{maru:ref} is
occasionally put in the beam to measure the kinematics of the pions
after their production and subsequent focusing. 
After the decay volume,
an iron and concrete beam dump stops essentially all charged
particles except muons with an energy greater than 5.5 GeV. 
Downstream of the dump there is 
a ``muon monitor'' (MUMON)\cite{maru:ref} consisting of a
segmented ionization chamber and an array of silicon pad detectors.
This monitors the residual muons spill-by-spill to check beam centering
and muon yield.
The ionization chamber consists of 5 cm wide strips covering
roughly $2\!\times\!2\rm m^2$ area
with separate planes for horizontal and vertical read-out.
The 26 silicon pads, each of \mbox{$\rm\sim 10 cm^2$} transverse area, are
distributed through a $3\!\times\!3\rm m^2$ area.

      Neutrino interactions near the production site are measured by
a set of detectors with complementary abilities. The detectors are located
approximately 300 m from the pion production target with 
approximately 70 m of earth 
                             eliminating virtually all prompt
beam products other than neutrinos.
A one kiloton water \Cherenkov detector (1kt) uses the same technology and
analysis algorithms as the far detector, \SuperK~(SK).
It has 680  20''  photomultiplier tubes (PMTs) on a 70 cm grid lining
a $\rm 8.6 m$ diameter, $\rm 8.6 m$ high cylinder.
The PMTs themselves and their arrangement are the same as in SK,
giving the same fractional coverage by photo-cathode (40\%).
A scintillating fiber detector (SciFi)\cite{Suzuki} with a 6 ton 
water target has tracking capability, 
and allows discrimination between different types of interactions
such as quasi-elastic or inelastic.
Downstream of the SciFi  there is a lead glass array 
for tagging electromagnetic showers.
A muon range detector (MRD)\cite{MUC:ref}, measures the 
energy, angle, and production point of muons 
from charged current (CC) $\nu_{\mu}$ interactions.  
It covers $7.6\!\times\!7.6\mathrm m^2$ transverse
area with four 10-cm iron plates, followed by 
eight 20-cm iron plates, all interleaved with drift tubes. 
The total mass is 915 tons.
The 6632 drift tubes each has $\rm 5\!\times\!7 cm^2$ cross section,
and are arranged in horizontal and vertical read-out planes.

      The far detector for K2K is the 
SK detector located in the Kamioka Observatory, Institute for Cosmic
Ray Research (ICRR), University of Tokyo, 
which has been taking data since 1996.
Its performance and results have been documented in the
literature\cite{SKatm:ref}.
Event selection for this detector uses timing synchronization with
the KEK-PS via the Global Positioning System (GPS)\cite{GPS:ref}.

      The beam-line was aligned by GPS position survey\cite{align:ref}.
The precision of this survey for the line from the target to the far
detector is better than 0.01 mr and the
construction precision for the near site alignment
is better than 0.1 mr.
The predicted neutrino spectrum at 250 km
is approximately the same over nearly 1 km,
giving a required accuracy of $\rm 3 mr$ for the pointing of the beam.

 The steering and the monitoring of the $\nu_{\mu}$ beam 
are carried out based on the MUMON measurement.
The $\nu_{\mu}$ and muons in the decay volume originate from the
same pion decays, so the measured center position of 
the muon profile 
is correlated with the $\nu_{\mu}$  beam direction.
The r.m.s. muon profile width is $\rm\sim 90 cm$ for both the horizontal
and vertical directions.
These measurements are used in tuning the beam direction
at the start of every beam period, usually once per month.
During beam tuning, the primary proton position
at the target is adjusted so that the muon profile is centered on 
the SK direction within 0.1 mr.
The center of the muon profile is measured to be stable
spill-by-spill within $\rm\pm 1 mr$ in each dimension
throughout the whole experimental period.
The stability of the muon yield is directly related to the stability
of the horn-focused $\nu_{\mu}$ beam.
The yield normalized to proton intensity measured by a current transformer
is stable spill-by-spill within a measurement uncertainty of $2\%$.

The characteristics and stability of the $\nu_{\mu}$ beam itself are
directly monitored at the near site
by the MRD using $\nu_{\mu}$ interactions with iron.
The large transverse area of MRD makes it possible to measure 
the beam direction and width (profile),
and the large mass makes it possible to measure the time stability 
of the $\nu_\mu$ 
event rate, profile and spectrum.
The location of the center of the profile gives the beam direction.
For the profile measurement, the starting point of a reconstructed muon track 
is regarded as the vertex of the $\nu_{\mu}$-iron interaction.  
The vertex distribution is corrected for geometrical acceptance 
to obtain the beam profile.  
The $\nu_{\mu}$ direction 
has been stable within $\pm1$ mr throughout all data-taking periods.  
The measured rate of $\nu_{\mu}$-iron events  is 
typically 0.05 events/spill and can be monitored on a daily
basis with good statistics. 
The event rate normalized to proton intensity is stable within
the statistical error of the MRD measurement.
The muon energy and angular distributions are also 
continuously monitored.  They show no change as a function of time,  
implying the $\nu_\mu$ energy spectrum is stable throughout this period.

      Comparison between expected and observed numbers of $\nu_{\mu}$  
events at the far site is done with this knowledge of measured
beam stability.
      To predict the $\nu_{\mu}$ beam characteristics at the far site, 
a normalization measurement at the near site and the extrapolation of the
information from near to far are necessary.
For the rate normalization, the 1kt is used 
so that any detector or analysis bias is suppressed.
For the extrapolation, the beam simulation is used, which is validated 
by the PIMON measurement as described below.
This simulation is based on GEANT\cite{GEANT:ref} with detailed 
description of 
materials and magnetic fields in the target region and decay volume.
It uses as input a measurement of the primary beam profile at the target.
Primary proton interactions on aluminum are modeled with
a parameterization of hadron production
data\cite{hadr:ref}.
Other hadronic interactions are treated by 
GEANT-CALOR\cite{GCALOR:ref}.

\begin{figure}[h!]
 \centerline{\psfig{file=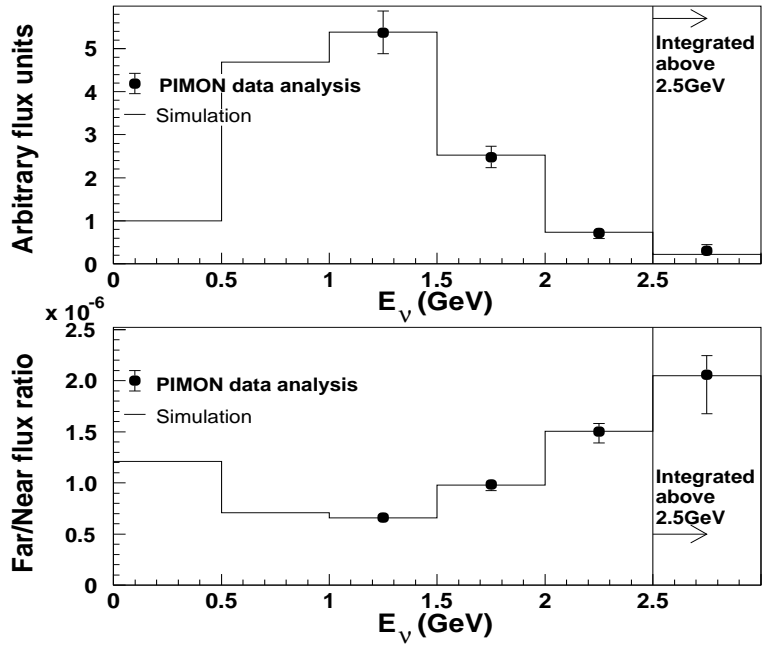,width=\figwid,height=0.85\figwid}}
   \caption{
             The top figure shows the $\nu_\mu$ energy spectral shape 
             at the near site and the bottom figure shows the far 
             to near $\nu_{\mu}$ flux ratio. 
             The histograms are from the beam simulation results. 
             The data points are derived from the PIMON measurement. 
             For the top figure, 
             normalization is done by area above 1 GeV.
         The vertical error bars for the data points reflect the
        total uncertainty.
        }
  \label{fig:beammon}
\end{figure}

Once the kinematic distribution 
of pions after production and focusing is known, it is possible
to predict the $\nu_{\mu}$ spectrum at any distance from the source.
The pion momentum and angular distribution is measured by the PIMON.
\Cherenkov photons generated by pions in the detector are focused
by a
pie-shaped 
spherical mirror to an array of 
PMTs of 8 mm
effective diameter in its focal plane.
Photons emitted by particles with the same velocity and 
incident angle with respect to the mirror arrive at 
the same position in the  focal plane.
This is  independent of  photon incident position with 
respect to the mirror.
The photon distribution on the PMT array is a superposition of slices 
of the \Cherenkov rings from particles of various velocities and angles. 
Measurements are made at seven indices of refraction, 
controlled by gas pressure, to give additional information on the
velocity distribution of the 
particles.
The indices of refraction are chosen so that the corresponding pion
momentum interval is 400 MeV/c.
The pion two-dimensional distribution of momentum versus angle is 
derived by unfolding the photon distribution data with various 
indices of refraction. 
To avoid background from 12 GeV primary protons,
the index of refraction
is adjusted below the \Cherenkov threshold of 12 GeV 
protons $(1-\beta\approx\!2.6\times 10^{-3})$.
As a consequence, analysis is done for $\rm P_\pi\ge 2 GeV/c$,
giving the $\nu_\mu$ energy spectral shape  above 1 GeV.
Figure~\ref{fig:beammon} shows the inferred
$\nu_\mu$ energy spectral shape at the near site and  
the far to near $\nu_{\mu}$ 
flux ratio along with the beam simulation result.
The beam simulation is well validated by the PIMON measurement
without any tuning.

The $\nu_\mu$  interaction rate at the near site is measured in the
1kt by detecting \Cherenkov light emitted from produced charged particles. 
The vertices and directions of \Cherenkov rings are 
reconstructed with the same methods as in SK\cite{subGeV:ref}.
In the 1kt, however, an analog sum of signals from all PMTs is recorded
by an FADC to select beam spills with only one event.
The definition of events per spill is the number of 
peaks of the FADC signal with $>\!1000$ collected photo-electrons (p.e.)
 ($\rm\approx\!100 MeV$ deposited energy). 
The average number of events observed in the full detector
is about 0.2/spill including entering background events
due to cosmic rays or $\nu_\mu$ induced muons from upstream
(about 1.5\% of events in the fiducial volume defined below). 
Thus $\ge\!2$ events occur in about 10\% of those spills
with $\ge\!1$ event. 
Neutrino interactions are selected by requiring: 
a) There is no detector activity in the $1.2 \rm\mu{s}$ preceding 
   the beam spill.
b) Only a single event is observed in the FADC peak search for that spill. 
c) The reconstructed vertex is inside the 25 t fiducial volume defined by
a 2 m radius, 2 m long cylinder along the beam axis,
in the upstream side of the detector.
%
The detection efficiency of the 1kt detector for detecting 
CC interactions is 87\% and for neutral current inelastic ($\rm NC_{inel}$) 
interactions 
it is 55\%. The expected ratio of CC to $\rm NC_{inel}$ gives 
an overall efficiency
of 72\%.
The main source of inefficiency is 
the 1000 p.e. threshold of the peak search in the FADC signal. 
The neutrino interaction simulator used for efficiency calculations 
is the same as that used in all SK analyses.
Finally the number of $\nu_\mu$ events in 1kt is corrected 
for spills with multiple events.
The average $\nu_\mu$ event rate per proton hitting the target is
$3.2\times 10^{-15}$.
Correcting for efficiencies, relative target masses, and detector live times,
the expected $\nu_\mu$ signal at the far site is estimated
by applying the extrapolation 
from the experimentally validated beam simulation.

Accelerator-produced neutrino interactions at the far site, SK, are 
selected by comparing two Universal Time Coordinated (UTC)
time stamps from the GPS system,
$T_{\mathrm{KEK}}$ for the KEK-PS beam spill start time 
and $T_{\mathrm SK}$ for the SK trigger time.
The time difference between two UTC time stamps, 
$\Delta(T)\equiv T_{\mathrm{SK}} - T_{\mathrm{KEK}} - TOF$ where 
$TOF$ is the time of flight of a neutrino,
should be distributed around the interval from
$0{\rm\ to}\ 1.1 \mathrm\mu{s}$
to match the width of the beam spill of the KEK-PS. 
Since the measured uncertainty of the synchronization accuracy
for the two sites is $\rm<200 ns,$
beam-induced events are selected in a $\rm 1.5 \mu{s}$ window.
\begin{figure}[htbp]
  \centerline{\psfig{file=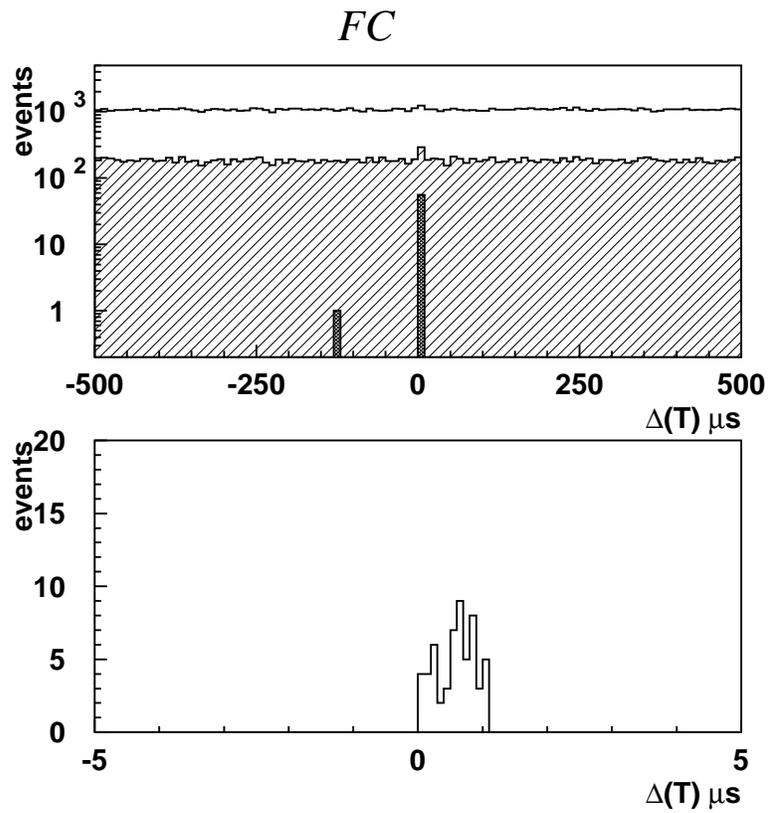,width=\figwid}}
  \caption{Far site timing analysis:
    (top) $\Delta(T)$ distribution over a $\pm\!500 \rm\mu{s}$ window for 
    events remaining after cuts a) (unfilled histogram)
                           and  b) (hatched),
                           and  after all cuts except timing (shaded).
    The cuts are described in the text.
    (bottom)
    $\Delta(T)$  for a $\rm\pm 5 \mu{s}$  window for FC events with 
    vertices in the fiducial volume.
}
\label{fig:tdif}
\end{figure}
Data reduction similar to that used in atmospheric neutrino 
analyses at SK\cite{subGeV:ref,multi:ref} is applied 
to select fully contained (FC) neutrino interactions. 
The criteria are:
a) There is no detector activity within 30 $\mu$s before the event. 
b) The total collected p.e.~in a 300 ns time window is $>200$
 ($\rm\approx\!20 MeV$ deposited energy).  
c) The number of PMTs in the largest hit cluster 
in the outer-detector is $<10$.
d) The deposited energy is $>30$ MeV.
Finally, a fiducial cut is applied accepting only events with fitted 
vertices inside the same $22.5\rm kt$ volume used for SK atmospheric
neutrino analysis.
The detection efficiency of SK is 93\% for CC interactions
and 68\% for $\rm NC_{inel}$ interactions, for a total of 79\%.
Similarly to the 1kt, the inefficiency is mainly due to the energy cut.
Figure~\ref{fig:tdif} shows the $\Delta(T)$ distribution 
at various stages of the reduction. 
A clear peak in time with the neutrino beam from the KEK-PS 
is observed in the analysis time window.

In order to compare the expected and observed $E_\nu$ spectrum,
the neutrino energy was reconstructed for the 1-ring $\mu$-like FC events 
in the SK fiducial volume assuming charged-current quasi-elastic interaction
($\nu+n\rightarrow \mu+p$).
For this interaction, the initial neutrino energy can be 
calculated from the momentum($p_\mu$) and angle($\theta)$ of 
the outgoing muons as
\begin{equation}
E_\nu=\frac{m_N E_\mu - m^2_\mu/2}{m_N-E_\mu+p_\mu \cos \theta_\mu},
\end{equation}
where $m_N$ and $m_\mu$ are masses of nucleon and muon, respectively.
It should be noted that for non-quasi-elastic interaction events,
the neutrino energy is underestimated.
The expected spectrum was made from the beam and neutrino interaction MC
simulation.
Overall normalization was done using the event rate measurement
at the 1kt.

\section{Results}

Fifty-six FC events are observed in the fiducial volume.
The $1.5\rm \mu{s}$ selection window gives an expected
background from atmospheric
neutrino interactions of order $10^{-3}$ events. 

The systematic uncertainty of the 1kt measurement is 4\%, for which the leading
terms are due to vertex fitting and its effect on the fiducial cut.
Other sources such as energy scale uncertainty and the treatment of 
multiple events in a spill are relatively small.
The statistical uncertainty of the 1kt measurement is $<1\%$.
The systematic uncertainty of the extrapolation from the near site measurement
is estimated to be $^{+6}_{-7}\%$ based on the PIMON measurement
uncertainty and beam simulation uncertainty for low energy neutrinos.
The systematic uncertainty in the SK measurement is
$3\%$, mainly due to the fiducial cut.
The systematic uncertainty term coming from uncertainty in the neutrino 
energy spectrum and cross section is small due to cancellations.
The quadrature sum of all known uncertainties for the far site
event rate prediction is $^{+9}_{-10}\%.$
The resulting expectation is $80.6^{+7.3}_{-8.0}$ events
in the absence of neutrino oscillations.
The observed and expected numbers for various categories are summarized
in Table~\ref{tab:sknum}.
The number of \Cherenkov rings and particle identification are reconstructed 
by the same algorithms as those used at SK\cite{subGeV:ref}, and the
event category definitions are also the same.
The expected number of events for analyses of CC
interactions in the other near detectors are
$87^{+13}_{-14}$ for MRD (iron target) and 
$88^{+11}_{-12}$ for SciFi (water target)\cite{K2Kfull:ref},
each of which is consistent with the value based on 1kt analysis.

 \begin{table}[htbp]
\begin{center}
\caption[Summary]
 {Summary of the observed and expected numbers of 
  FC events in the fiducial volume at the far site.}   
 \begin{tabular}{ccc}
\hline
\hline
    Event Category & Observed   &  Expected \\
 \hline
 \hline
    Single Ring $\mu$-like       &  30   & 44.0$\pm$6.8  \\
    Single Ring $e$-like         &  2    & 4.4$\pm$1.7  \\
    Multi Ring                   &  24   & 32.2$\pm$5.3  \\
 \hline
    TOTAL & 56    & $80.6^{+7.3}_{-8.0}$  \\
\hline
\hline
 \end{tabular}
 \label{tab:sknum}
\end{center}
 \end{table}

\begin{figure}[htbp]
  \centerline{\psfig{file=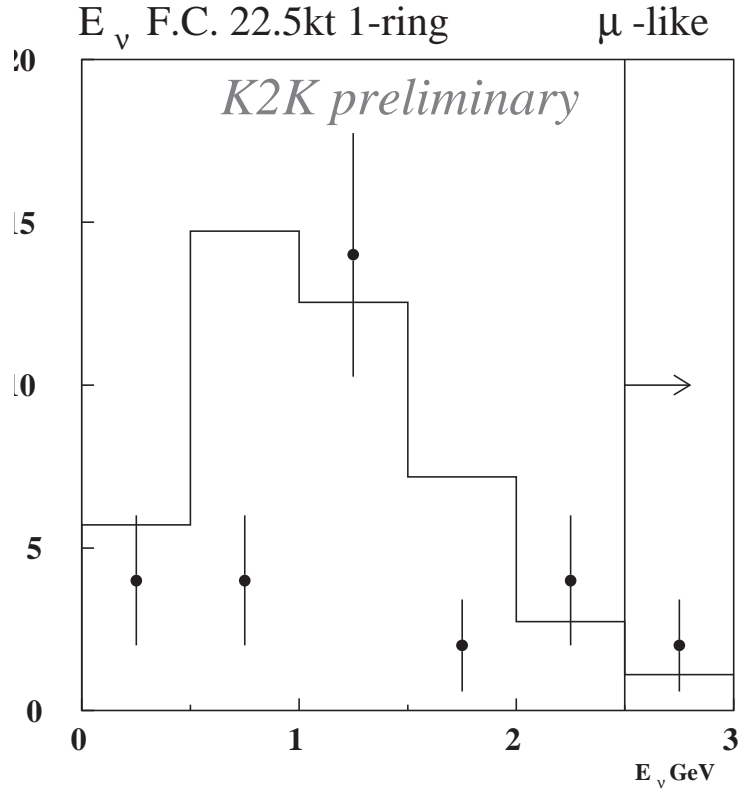,width=\figwid}}
  \caption{Spectrum of reconstructed $E_\nu$ for 
the 1-ring $\mu$-like FC events in the SK fiducial volume.
The histogram is the one expected from the beam and neutrino interaction MC.
}
\label{fig:enu}
\end{figure}

Figure~\ref{fig:enu} shows the reconstructed $E_\nu$ spectrum
for the 1-ring $\mu$-like FC events in the SK fiducial volume.
The histogram in the figure is the one expected from the beam and neutrino 
interaction MC.
The difference between the expected and observed spectrum
indicates the distortion due to the neutrino oscillation.

\section{Conclusion and Future Prospect}

  The crucial principles of     K2K            have been proven to work:
  The beam direction has been continuously monitored and controlled 
within $\rm\pm1 mr$ for several months duration by controlling the 
proton position on target with millimeter accuracy and operating the
horn magnet stably.
  Pion kinematics have successfully been measured {\em in situ\/} 
allowing prediction of the neutrino beam at the distance of 250 km
with six to seven percent accuracy.
  Events at the far site have been identified by means of GPS timing,
reducing backgrounds to a negligible level.
Fifty-six FC neutrino events have been observed where 
$80.6^{+7.3}_{-8.0}$ are expected.
The study of the spectral shape for the oscillation analysis is now on going.
In order to confirm the neutrino oscillation phenomena and to determine
the oscillation parameter precisely, 
the beam spectral shape, neutrino interaction and their errors
are being carefully studied.
For more complete study of the beam spectrum and neutrino interaction,
a new fine-grained($\sim$2cm) detector will be installed in 2003.
It consists of 3m long plastic scintillator bars and wavelength shifter fibers.
The experiment expects to accumulate $10^{20}$ protons on
target, providing sufficient statistics to study neutrino oscillations by
spectral analysis for $\nu_{\mu}$ disappearance.

\section*{Acknowledgments}
We thank the KEK and ICRR Directors for their strong support
and encouragement.
K2K is made possible by the inventiveness and the diligent
efforts of the KEK-PS machine group.
We gratefully acknowledge the cooperation of the
Kamioka Mining and Smelting Company.
This work has been supported by the
Ministry of Education, Culture, Sports,
Science and Technology, Government of Japan 
and its grants for Scientific Research, 
the Japan Society for Promotion of Science, 
the U.S. Department of Energy, the Korea
Research Foundation, and the Korea Science and Engineering Foundation.

\section*{References}
\def\etal{{\em et al.}}

\end{document}